\newcommand{\ale}{\ \raisebox{-.3ex}{$\stackrel{<}{\scriptstyle \sim}$}\ }
\newcommand{\age}{\ \raisebox{-.3ex}{$\stackrel{>}{\scriptstyle \sim}$}\ }

\documentstyle{mn}

\input psfig

\title[Disc evolution in T Tauri stars]
	{Accretion disc evolution in single and binary T Tauri stars}
	
\author[P.J. Armitage, C.J. Clarke and C.A. Tout]{P.J. Armitage$^1$, 
	C.J. Clarke$^2$ and C.A. Tout$^{2,3}$ \\
	$^1$ Canadian Institute for Theoretical Astrophysics, McLennan 
	Labs, 60 St George Street, Toronto, Ontario M5S 3H8, Canada \\
 	$^2$ Institute of Astronomy, Madingley Road, Cambridge, CB3 0HA \\
 	$^3$ Department of Mathematics, Monash University, Clayton, 
 	Victoria 3168, Australia}	
 	
\begin{document}

\maketitle

\begin{abstract}
We present theoretical models for the evolution of T Tauri stars 
surrounded by circumstellar discs. The models include the effects 
of pre-main-sequence stellar and time dependent disc evolution, and 
incorporate the effects of stellar magnetic fields acting on the inner
disc. For single stars, consistency with observations in Taurus-Auriga
demands that disc dispersal occurs rapidly, on much less than the 
viscous timescale of the disc, at roughly the epoch when heating 
by stellar radiation first dominates over internal viscous dissipation.
Applying the models to close binaries, we find that because the initial 
conditions for discs in binaries are uncertain, studies of extreme mass ratio 
systems are required to provide a stringent test of theoretical disc evolution 
models. We also note that no correlation of the infra-red colours of T Tauri 
stars with their rotation rate is observed, in apparent contradiction
to the predictions of simple magnetospheric accretion models. 
\end{abstract}


\section{Introduction}

Observations of pre-main-sequence T Tauri stars suggest that
stellar magnetic fields are often important in the inner regions
of surrounding circumstellar discs. The field acts to provide
a link between the star and its disc, and the resulting torques
lead to truncation of the disc interior to some magnetospheric
radius $R_{\rm m}$ (see e.g. Pringle \& Rees 1972;
Ghosh \& Lamb 1979). This divides the accretion flow into two
regimes; an outer disc where inflow is slow and the surface
density of the disc material relatively large, and an inner evacuated
`hole' where the magnetic field is dominant and infall onto the
star proceeds at approximately the free-fall velocity. 

Theoretical work shows that stellar magnetic fields of the order 
of $\sim 1\,{\rm kG}$ are a prerequisite for this mode of
accretion (K\"onigl 1991). Magnetic fields of this
strength have been directly detected in a few weak-line
T Tauri stars (WTTS - Basri, Marcy \& Valenti 1992; Guenther \&
Emerson 1995, 1996), while the high X-ray flux (Montmerle 1992;
Neuh\"auser et al. 1995) and flaring activity (Preibisch, Zinnecker \&
Schmitt 1993) seen in the general T Tauri population suggest that strong
magnetic fields are common. Several {\em consequences} of
magnetospheric accretion are also observed, including
mass infall at free fall velocities (Calvet \& Hartmann 1992;
Edwards et al. 1994; Martin 1997; Johns-Krull \& Hatzes 1997), 
the presence of hotspots in classical 
T Tauri stars (CTTS - Bouvier et al. 1993), and infra-red colours
that are too red to match disc models that extend to the
stellar surface (Bertout, Basri \& Bouvier 1988). 

Theoretical models that aim to describe the star-disc interaction
are abundant (e.g. Ghosh \& Lamb 1979; Campbell 1992; Yi 1994;
Lynden-Bell \& Boily 1994; Wang 1995, 1996; Lovelace, Romanova \& 
Bisnovatyi-Kogan 1995; Ostriker \& Shu 1995). There are important 
differences between these models, but a generic prediction
is that the magnetospheric radius should lie close to the 
corotation radius ($R_{\rm c}$, where the the disc material 
is stationary in the rotating stellar frame), if the magnetic 
linkage is to be effective in regulating the star's angular 
momentum (Wang 1995; Armitage \& Clarke 1996). 

In recent studies, Kenyon, Yi \& Hartmann (1996) and Meyer, 
Calvet \& Hillenbrand (1997), compared the infra-red colours of 
a sample of classical T Tauri stars with the predictions of several 
steady-state magnetic and non-magnetic disc models. They concluded 
that the magnetically disrupted discs provided a better 
fit to the observed colours, and were able to place some constraints
on the location of the magnetosphere relative to corotation. 

In this paper, we employ time-dependent theoretical models 
to investigate how magnetically truncated discs evolve with 
time. Because the viscous timescale in the region where 
infrared emission originates ($R < 1 \ {\rm a.u.}$) is small, 
we expect that steady-state models such as those used previously 
will suffice for calculating the infrared colours (assuming 
that the inner regions are stable). However the use of a 
time-dependent approach permits us to study how rapidly the 
system evolves in, for example, the infrared colour-colour plane. 

Section 2 describes the numerical models used in this work, which 
are applied in Section 3 to follow the evolution in the infra-red colour-colour 
diagram of discs surrounding magnetic T Tauri stars. For a limited 
number of systems additional information, in the form of measured 
stellar rotation periods, is available, and we consider the
constraints this imposes in Section 4. Section 5 applies our models 
to discs in close binary systems. Section 6 summarises our conclusions. 

\section{Numerical modelling}

The numerical models employed in this work are derived from
those used previously by Armitage \& Clarke (1996) for modelling
T Tauri systems with discs. We follow the evolution
of a theoretical pre-main-sequence stellar model surrounded by
a geometrically thin $\alpha$-prescription (Shakura \& Sunyaev
1973) accretion disc. A treatment that fully coupled the 
stellar evolution calculation to the disc would be 
essential for the early stages of accretion, here we only
consider accretion rates low enough that the stellar evolution 
and accretion problems can be considered separately.

\subsection{Stellar model}

\begin{figure}
 \psfig{figure=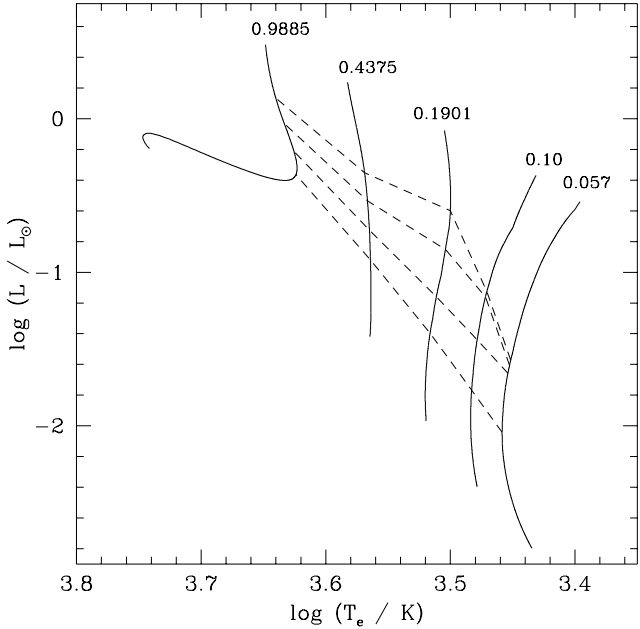,width=3.5truein,height=3.5truein}
 \vspace{0.0truein}
 \caption{HR diagram for the stellar models used in this paper,
 	  with stellar masses $M_* =$ 0.9885, 0.4375, 0.1901, 0.10
 	  and 0.057 $M_\odot$. Tracks are plotted started
 	  (arbitrarily) where $R_* = 3 \ R_\odot$, with
 	  isochrones shown relative to this time at 
 	  $\Delta t = 1, 2, 4, 8$ Myr.}
 \label{figure1}
\end{figure} 

\begin{figure}
 \psfig{figure=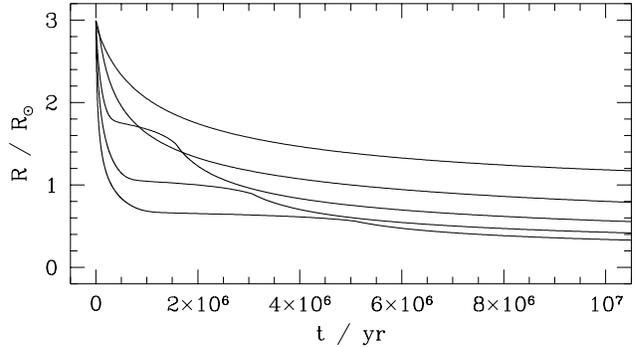,width=3.5truein,height=3.5truein}
 \vspace{-1.5truein}
 \caption{Time evolution of the stellar radii, for models with
 	  stellar masses (from top downwards)
 	  $M_* =$ 0.9885, 0.4375, 0.1901, 0.10
 	  and 0.057 $M_\odot$.}
 \label{figure2}
\end{figure}

Figure \ref{figure1} shows the Hertzprung-Russell diagram for the
theoretical pre-main-sequence models employed in this paper.
We construct our stellar models using the most recent version of the
Eggleton evolution program (Eggleton 1971, 1972, 1973).  The equation of
state,
which includes molecular hydrogen, pressure ionization and coulomb
interactions, is discussed by Pols et al. (1995).
The initial composition is supposed to be uniform with a hydrogen
abundance $X = 0.7$, helium $Y = 0.28$, deuterium $X_{\rm D} =
3.5\times 10^{-5}$ and metals $Z = 0.02$ with the meteoritic mixture
determined by Anders and Grevesse (1989).  Hydrogen burning is allowed
by the pp chain and the CNO cycles.  Deuterium burning is explicitly
included at temperatures too low for the pp chain.  Once the pp chain
is active hydrogen is assumed to burn to He$^4$ via deuterium and
He$^3$ in equilibrium.  The burning of He$^3$ is not explicitly followed.
Opacity tables are
those calculated by Iglesias, Rogers and Wilson (1992) and Alexander
and Ferguson (1994).  An Eddington approximation (Woolley and Stibbs
1953) is used for the surface boundary conditions at an optical depth
of $\tau = 2/3$ so that low-temperature atmospheres, in which
convection extends out as far as $\tau \approx 0.01$ (Baraffe et
al. 1995), are not modelled perfectly.  However the effect on
observable quantities (see Kroupa and Tout 1997) is not significant
in this work.

The tracks in figure~1 are for four pre-main-sequence stars of masses
0.9885, 0.4375, 0.1901, 0.1 and $0.057\,M_\odot$ shown from a radius
of $3\,R_\odot$ until they begin hydrogen burning at the main
sequence.  The most massive model establishes a radiative core well
above the main sequence and so follows a Henyey track in its last
stages of contraction.  The other models remain fully convective
although the $0.4375\,M_\odot$ star will develop a radiative core on
the main sequence.
The actual starting point for all of our models was an accreting
protostellar track in which material of the zero-age composition and
with the surface state is accreted at a rate of $10^{-5}\,M_\odot\,{\rm
yr}^{-1}$.  Accretion was interrupted when the various total masses
were reached and the stars allowed to evolve freely with normal
photospheric boundary conditions and constant total mass.  All these
models have relaxed from their accreting state by the time they reach
$3\,R_\odot$ and are by that time descending Hayashi tracks as typical
pre-main-sequence stars.  Figure~2 illustrates the evolution of radius
with time for each of these stars.  The three lower-mass objects show
a plateau lasting between one and five million years.  This corresponds
to the delay in shrinkage while the stars burn their primordial
deuterium.  The duration of this plateau will depend on the actual
value assumed for the primordial deuterium abundance.  The two more
massive stars have already burnt all their deuterium by the time they
have shrunk to $3\,R_\odot$ because their cores are correspondingly
hotter.

The early accretion history of protostars has been considered 
by Stahler (1988), and by Hartmann, Kenyon \& Cassen (1997). How the 
accretion influences pre-main-sequence evolution depends upon whether 
accretion at early times is via a magnetospheric or boundary layer 
mode, and on the amount of mass gained during high accretion rate outbursts.
Here we are primarily interested in the later phases of accretion, 
for which purpose a rather arbitrary assignment of $t = 0$ suffices.
We choose the time at which $R_* = 3\,R_\odot$ because it
corresponds to roughly to the upper limit
on the inferred stellar radii of CTTS (Hartigan, Edwards \& Ghandour
1995), and the radii used for models of stars suffering FU Orionis 
outbursts at higher accretion rates than those modelled here (Bell \&
Lin 1994).

\subsection{Disc model}

We consider a geometrically thin non-self-gravitating disc evolving 
under the influence of viscous and stellar magnetic torques.
The evolution of the disc surface density distribution $\Sigma(R,t)$ is
described by (e.g. Livio \& Pringle 1992),
\begin{equation}
 { {\partial \Sigma} \over {\partial t} } = 
 { 3 \over R } { \partial \over {\partial R} } \left[
 R^{1/2} { \partial \over {\partial R} } \left(
 \nu \Sigma R^{1/2} \right) \right] +
 { 1 \over R } { \partial \over {\partial R} } \left[
 { {B_z B_\phi R^{5/2}} \over {\pi \sqrt{GM} } } \right] .
\label{eq_evolve}
\end{equation} 
Here $\nu$ is the kinematic viscosity, $M$ the stellar mass,
and $B_z$ and $B_\phi$ the vertical and azimuthal components 
of the stellar magnetic field at the surface of the disc. 
We make the standard assumption that the vertical
field $B_z$ has a dipolar fall-off with radius at the magnetospheric
radius (generally at several $R_*$), and use the expression for
the azimuthal field given by Wang (1995),
\begin{eqnarray}
 { B_\phi \over B_z } = { {\Omega(R) - \Omega_*} \over \Omega(R) } \ \
 \Omega \ge \Omega_* \nonumber \\ 
 { B_\phi \over B_z } = { {\Omega(R) - \Omega_*} \over \Omega_* } \ \
 \Omega < \Omega_* ,
\label{eq_torque}
\end{eqnarray} 
where $\Omega(R)$ is the Keplerian velocity at radius $R$ in the
disc and $\Omega_*$ the angular velocity of the stellar surface.
In protostellar discs where the Shakura-Sunyaev $\alpha$ parameter 
is probably $\ll 1$, this expression is likely to be valid if 
rapid reconnection in the corona limits the growth of $B_\phi$.
Different field geometries arise if $\vert B_\phi / B_z \vert$ is able to 
grow to a large value (Bardou \& Heyvaerts 1996).

The kinematic viscosity is described using solutions to the
vertically averaged equations for thin discs given in Faulkner,
Lin \& Papaloizou (1983). The analytic opacity fits used by
Bell \& Lin (1994) are used, so that in any one opacity regime
the solution for $\nu(\alpha,R,\Sigma)$ is a power-law in 
surface density, radius, and $\alpha$
parameter. Transitions to different dominant opacities
occur as the disc becomes cooler at greater radii and later times.
At an accretion rate of $\dot{M} = 10^{-8} \ M_\odot \ {\rm yr}^{-1}$,
and radii between 5 and 80 $R_\odot$, these solutions agree with
the detailed vertical structure integrations presented by
Bell \& Lin (1994), using the same opacities, to within
typically 20-30 \%. 

At low accretion rates, the thermal structure of the disc
will be determined primarily by the heating from the central
star, rather than viscous dissipation within the disc. When the
accretion rate falls low enough that this occurs, we keep 
$\nu(R)$ constant for the remainder of the evolution.

\subsection{Spectral energy distributions}

We calculate the disc spectral energy distribution assuming 
that each annulus in the disc radiates locally the energy
input from viscous heating and reprocessing of stellar radiation.
For models including a stellar magnetic field the work done
on the disc material by magnetic torques is also included. The
total heating of one side of the disc is then,
\begin{equation}
 Q^+ = Q^+_{\nu} + Q^+_{\rm B} + Q^+_{\rm p},
\label{2.3.1}
\end{equation}
where the viscous term is the usual expression,
\begin{equation}
 Q^+_{\nu} = {9 \over 8} \nu \Sigma \Omega^2,
\label{2.3.2}
\end{equation}
and the magnetic contribution is given by
\begin{equation}
 Q^+_{\rm B} = {1 \over 2} (B_z B_\phi)_{z=H} R^2 \vert \Omega - \Omega_*
 \vert .
\label{2.3.3}
\end{equation}  
The component $Q_{\rm p}^{+}$ due to reprocessing of stellar 
radiation is computed using the standard expressions given 
by Kenyon \& Hartmann (1987; see also Adams \& Shu 1986). We
assume a flat disc, a stellar spectrum given by a Planck 
function $B_\nu (T_{\rm e})$, and allow for the partial 
absorption when the disc starts to become optically thin.

Treating the disc as an isothermal slab, the energy loss rate per unit 
area through one side of the disc $Q^{-}$ is,
\begin{equation} 
 Q^{-} = \pi B_\nu (T_{\rm disc}) \left[ 1 - 2 E_3 (\tau_0) \right]
\label{2.3.7}
\end{equation}
where,
\begin{equation} 
 E_3 (\tau_0) = \tau_0^2 \int_{\tau_0}^{\infty} { {\exp^{-t} {\rm d}t} 
 \over {t^3} },
\label{2.3.8}
\end{equation}
$\tau_0 (\nu) = \kappa(\nu) \Sigma$ is the optical depth normal 
to the disk plane, and $T_{\rm disc}$ is determined by the requirement 
that $Q^{+} = Q^{-}$.

The dust opacity $\kappa(\nu)$ is that used by
Men\' shchikov \& Henning (1997) in their modelling of the source
L1551 IRS 5, and incorporates a mixture of carbon and silicate components.
Dust opacities vary considerably with the adopted grain model (Men\' shchikov,
private communication, see also figure 15 of Men\' shchikov \& Henning 1997),
and this will modify the SED as the disc becomes optically thin.
Additionally in the inner regions of the disc, where $T \age 1500$ K,
the dust is destroyed and the remaining opacity, owing
to molecules, is greatly reduced (Calvet, Pati\~no \& Magris 1991; Alexander,
Johnston \& Rypma 1983). This is not included in our current 
calculations. However, at low accretion rates, when the temperature
distribution in the disc is dominated by the reprocessing of stellar 
radiation, we expect this to be an important complication only 
in the innermost regions of the disc. For example, for a star 
of effective temperature $T_* = 4000$ K surrounded by a flat 
disc, the temperature falls below 1500 K within $3 R_*$. 
In our evolutionary models the magnetosphere generally extends 
to at least this radius at late times, and so we do not expect 
large changes in our results if dust destruction was treated
consistently.

For each model, we compute the spectral energy distribution (SED) 
for two system inclinations, a face-on disc system and one inclined 
at 60$^{\circ}$. Infra-red colours are then computed using the 
observed colours of main-sequence stars of given effective temperature 
(Kenyon \& Hartmann, 1995), to which are added the contribution from 
the disc using the zero points in the UKIRT system (e.g. as quoted in 
the HST {\em NICMOS} reference manual). 

\section{Colour evolution of magnetospheric accretion models}

\subsection{Initial conditions}

The basic properties of T Tauri discs are now known observationally. 
Typical disc masses are $\sim 10^{-2} \ M_\odot$ (Osterloh \& Beckwith 1995), 
typical accretion rates are $\sim 10^{-8} \ M_\odot {\rm yr}^{-1}$ 
(Gullbring et al. 1998), and typical disc lifetimes are a few Myr 
(Strom 1995). We aim to set up initial conditions that are simultaneously 
compatible with these observations, while noting that observationally there 
is considerable scatter in all of these quantities. We especially caution 
as to the possible systematic errors in estimating ages from pre-main-sequence 
tracks (Tout, Livio \& Bonnell 1998). 

For our magnetic models we use the $M=0.4375 \ M_\odot$ stellar model, 
and an initial steady-state disc with an accretion rate of 
$3 \times 10^{-7} \ M_\odot {\rm yr}^{-1}$, at the upper end of the 
T Tauri accretion rate distribution. The initial disc mass is taken 
as $0.1 \ M_*$. With these parameters, experimentation showed that 
an $\alpha$ of 0.02 produced sensible values for the disc lifetime 
(in the Myr range) and initial radius ($\sim 30$ a.u., though the 
disc would expand as it evolves). This value of $\alpha$ is also 
consistent with the findings of Hartmann et al. (1998), who considered
the evolution of rather different self-similar disc models.
These runs have a zero-torque outer 
boundary condition imposed at 30 a.u., and use 800 radial mesh points. 

The models also require specification of the stellar magnetic field and 
stellar rotation period. We parameterise the stellar
field as,
\begin{equation}
B_* = B_0 \left( P_* \over {4 \,{\rm d}} \right)^{-1},
\label{eq_B}
\end{equation}
and take $B_0$ between 250\,G and 2\,kG. For the stellar rotation rate,
we assume that $P_*$ varies smoothly between
an initial spin period $P_{\rm i} = 7 \,{\rm d}$, and a
final spin period $P_{\rm f} = 3.5 \,{\rm d}$, consistent
with observations (Bouvier et al. 1995). The transition
is smoothed according to,
\begin{equation}
 P_* = P_{\rm f} + { {(P_{\rm i} - P_{\rm f})} \over 
 {1 + (t / t_{\rm WTTS})^2} },
\label{eq_Poft}
\end{equation}
where the time taken to reach a WTTS spin period, $t_{\rm WTTS}$,
is taken to be 3 Myr. 

\subsection{Results}

\begin{figure}
 \psfig{figure=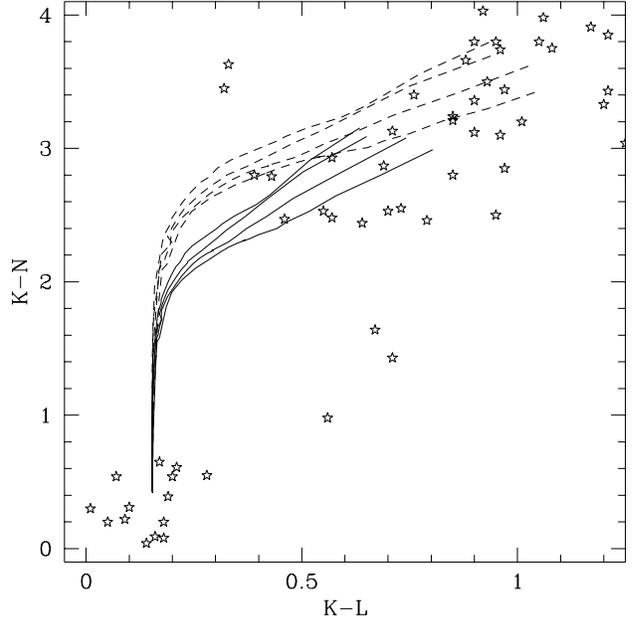,width=3.5truein,height=3.5truein}
 \caption{Tracks in the K-L, K-N plane for magnetically truncated
 	  disc models surrounding an $M_* = 0.4375 \ M_\odot$
 	  star, with $B_0 =$ 250, 500, 1000, and 2000 G
 	  respectively (higher $B_0$ models are redder in K-N).
          The dashed lines are calculated for a face-on disc 
          system, the solid lines for a disk inclined at 60$^\circ$.
 	  Symbols are from the compilation of observations by Kenyon \&
 	  Hartmann (1995).}
 \label{figure3}
\end{figure}

\begin{figure}
 \psfig{figure=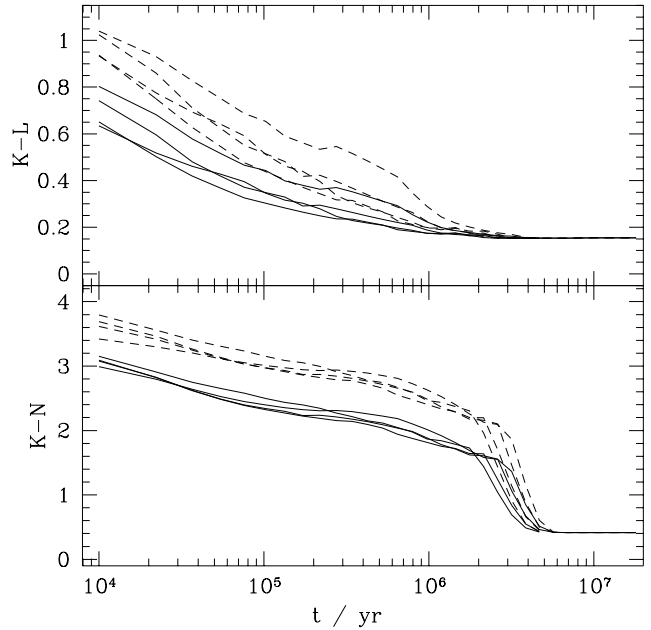,width=3.5truein,height=3.5truein}
 \vspace{0.0truein}
 \caption{Time evolution of the infra-red colours corresponding 
          to the tracks shown in Fig.~3. The dashed lines are 
          calculated for a face-on disc system, the solid lines 
          for a disk inclined at 60$^\circ$. Note that the 
          {\em absolute} time scaling is entirely determined 
          by the assumed value of $\alpha$, which is here 
          chosen as $\alpha = 0.02$ in order to roughly 
          match the observed lifetime of T Tauri discs.}
 \label{figure4}
\end{figure}

Figures \ref{figure3} and \ref{figure4} shows the model tracks in 
the (K-L)-(K-N) colour plane, together with evolution of the 
colours with time. Also plotted is the observed distribution of Class II 
and Class III sources in Taurus-Auriga. The data are taken from 
the compilation of Kenyon \& Hartmann (1995) and are not dereddened.
The stars comprising the sample have a fairly broad distribution 
of ages (Kenyon \& Hartmann 1995; see especially Fig.~16), with 
a typical age of $\sim 1$ Myr. A `gap' in K-N between $1 < (K-N) < 2$ is 
clearly seen, for this sample the {\em only} systems lying in the gap 
are three binaries without separate colour information for the two components.

For a given K-L, models with stronger stellar
magnetic fields are generally redder in K-N. Those models
with lower $B_0$, where for high accretion rates $R_{\rm m}$
is significantly smaller than $R_{\rm c}$, seem to fit the
envelope of the CTTS observations better than models with higher
stellar fields, as noted previously (Kenyon, Yi \& Hartmann 1996).

Our calculation of disc colours is crude -- discs are 
probably not flat, and their spectra in the infra-red is likely
to deviate substantially from a blackbody. Thus the tracks
taken by the models are more significant than the absolute match
with the data. Two phases of evolution occur. In the first phase,
the K-L colour declines smoothly to photospheric values, while the K-N
colour remains red. This first phase is driven by the decline in the accretion
rate, which increases the relative importance of magnetic torques
compared to the viscous torques in the inner disc, and
moves the magnetosphere outward towards corotation, destroying
the L flux from the disc and reducing the K-L colour. The N
flux, sampling cooler parts of the disc unaffected 
by the magnetosphere, remains strong. This longer wavelength
emission lasts until the inner disc becomes optically thin.
The inclusion of the magnetic torques does not assist in producing 
a sharp transition between disc-like and stellar K-N colours, 
which instead proceeds on the slow viscous timescale
of the outer disc (note that Fig.~\ref{figure4} is plotted 
on a logarithmic timescale, and thus the models predict 
that the amount of time spent in the unpopulated gap region 
is roughly equal to that spent in the allowed CTTS part of the 
diagram). The observation of a `gap' thus unequivocally 
requires disc clearing to take place on a timescale much shorter 
than expected from purely viscous disc processes, even if 
stellar magnetic fields are strong enough to rapidly 
evacuate the inner regions of the disc out to corotation.
Since mid-IR disc emission originates primarily from dust, 
this conclusion need not apply to the {\em gas} in the disc. 
A plausible model might envisage rapid dust agglomeration 
in the inner disc at the end of the CTTS phase, producing
the observational effects discussed above, while the gas
is dispersed on a more leisurely timescale. If this were
the case, we would expect there to be systems where signatures
of accretion were visible (e.g. strong H$\alpha$ emission), but 
which lacked disc emission in the infrared.

If magnetospheric accretion is commonplace, the models 
predict that the colours of T Tauri stars evolve into a region of
the colour-colour plane characterised by near-photospheric
K-L colour, and disc-like K-N colour. This occurs for all
values of the stellar magnetic field strong enough to 
disrupt the inner disc at CTTS accretion rates. Strikingly,
no systems are observed in this region of the diagram. Numerically, 
our model tracks move into this `empty' region of the diagram 
at accretion rates of $\dot{M} \sim 10^{-8} \ M_\odot {\rm yr}^{-1}$. 
This suggests either that discs are dissipated prior to stars reaching 
this stage (which would imply that passive reprocessing 
discs are likely to be rare), or that our models are missing 
significant extra sources of infrared flux that would shift 
the tracks redward into consistency with the data.
 The latter possibility cannot be ruled
out, though if the extra flux arises from a smaller magnetosphere 
(for example because the magnetic coupling is weaker than assumed
here), the implication would be that stellar angular momentum regulation 
via magnetic coupling to the disc would be very difficult to 
accomplish.

Since we have already emphasized that there is considerable 
uncertainty in our colour calculation, we note that the same 
conclusion (that reprocessing dominated discs appear to be rare) 
follows from noting that there is no apparent pile up of systems 
at the lower boundary of the CTTS colour distribution, as would 
be expected if there were a long-lived population of discs with 
very low accretion rates. 

\section{Rotation-colour correlation}

In the preceding Section we considered the infra-red colour distribution
of a large sample of T Tauri stars. For a limited 
subset of these, important additional information in the form
of photometric periods is available. If this also represents
the stellar spin period it provides a direct measure
of the location of corotation, which in some models is in
turn approximately equal to the magnetospheric radius.
More generally, if the magnetic coupling between the star and
its disc is to act to regulate the stellar angular momentum,
then the rapid dipolar fall-off of $B_z (R)$ implies that 
$R_{\rm m}$ must lie close to $R_{\rm c}$. This scenario then 
predicts that $R_{\rm m}$ should increase with increasing $P_*$, and
lead to a correlation between the stellar spin period and the 
infra-red colour. 

\begin{figure}
 \psfig{figure=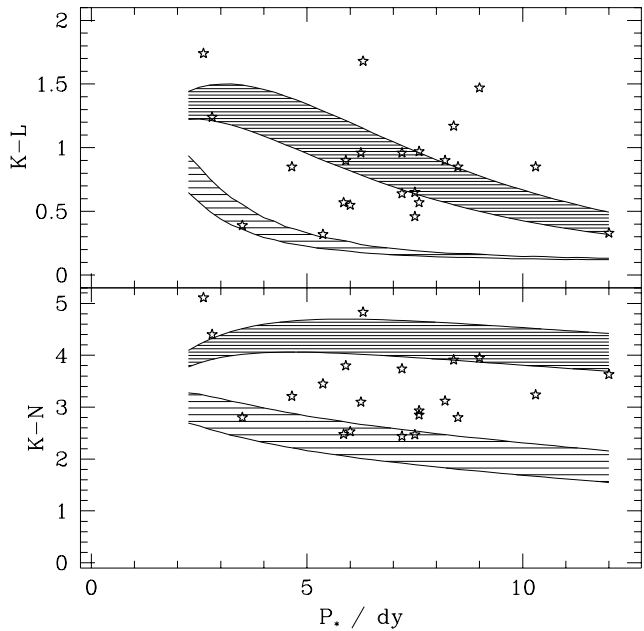,width=3.5truein,height=3.5truein}
 \vspace{0.0truein}
 \caption{K-L and K-N colours (from Kenyon \& Hartmann, 1995)
 	  for the CTTS with measured photometric periods $P_*$
 	  in the sample of Bouvier et al. (1995). The upper shded
 	  area shows the theoretical prediction for a magnetic
 	  disc model truncated at $R_{\rm m} = R_{\rm c}$, with
 	  $\dot{M} = 3 \times 10^{-7} \ M_\odot \ {\rm yr}^{-1}$, 
 	  viewed at angles between face-on and inclined by 60$^{\circ}$ 
 	  (other assumptions as stated in the text). The lower shaded
 	  area shows the predictions for a pure reprocessing 
 	  disc around the same star with $\dot{M} = 0$.}
 \label{figure5}
\end{figure} 

Figure \ref{figure5} shows the K-L and K-N colours of the 
CTTS with measured photometric periods in the sample compiled
by Bouvier et al. (1995). We make the standard assumption
that the photometric period traces the rotation rate of
the stellar surface and plot the colours as a function of $P_*$.
The sample comprises 22 systems, all of which have measured K-L 
colours in the compilation of Kenyon \& Hartmann (1995), and all
bar one of which have measured K-N colours.

We make no pretence that this sample is in any way complete
or unbiased. However the systems sampled
here do cover a broad span of stellar periods, from 2 to
12 days, which represents a factor of about 3 in the
implied radius of corotation. No trend
in the infra-red colours is seen with varying stellar
rotation period. This is obvious by eye, and a K-S test
(for example) readily confirms that the K-L colours for the 
sample split into `slow' and `fast' rotators are consistent
with these being drawn from a single distribution.

The Figure also shows the predictions of a steady-state version 
of the theoretical disc model described in Section 2
surrounding stars with varying rotation rates. We
assume $R_{\rm m}=R_{\rm c}$, and take for the stellar parameters
$M_*=M_\odot$, $R_*=2 \ R_\odot$, $B_* = 1\,$kG, and $T_* = 4400$ K.
The magnetic heating term is included. The hashed area shows
the range of colours expected for a disc model with an accretion rate of 
$3 \times 10^{-7} \ M_\odot {\rm yr}^{-1}$, viewed at 
different inclinations to the line of sight. Colours for a
reprocessing disc devoid of any internal source of luminosity are
also plotted. 

Experimentation with the models shows that judicious variation 
of $\dot{M}$ and $B_*$ can roughly reproduce the range of colours 
of these CTTS (though the reddest stars with long rotation 
periods would demand very high accretion rates).
 However, if $P_*$ is assumed to correlate well 
with $R_{\rm m}$, the models all predict a strong dependence 
of K-L colour with rotation period. This correlation is not 
seen in the current data. Note that K-N colour is not expected 
to vary greatly with $P_*$, as most of the 10 $\mu$m flux comes 
from greater radii and is unaffected by magnetospheres of this size.

The observations display a prominent scatter that could be
due to variations in accretion rate at a given rotation 
period or to factors not modelled in this analysis. 
However the dispersion, although large, would not
mask the expected trend if it were uncorrelated with
$P_*$. In particular, if we assume that the model 
with $\dot{M} = 3 \times 10^{-7} \ M_\odot {\rm yr}^{-1}$
defines the predicted theoretical trend, then the 
distribution of (K-L)$_{\rm data}$ - (K-L)$_{\rm model}$ 
does differ significantly between slow and fast rotators. 
Better samples and modelling of the colours
will be required to determine if this 
remains a problem for magnetospheric accretion 
models.   

\section{Disc evolution in binary systems}

A large fraction of T Tauri stars are members
of binary or multiple systems (Ghez 1996), and some
of these are close enough to influence
the evolution of circumstellar discs. We consider here
the evolution of discs in these moderately close 
($\sim 10 - 100 \ {\rm a.u.}$) binaries.

\subsection{Truncation radii}

Previous theoretical work has shown that the gravitational
influence of a binary system on circumstellar material is 
expected to be twofold: the individual circumstellar
discs will be truncated by the tidal effects of the
companion star and a gap will be cleared between
the binary and any surrounding material forming a 
circumbinary disc. The truncation and gap-clearing 
processes occur rapidly, on a dynamical timescale,
and so for our purposes the initial conditions for
discs in a binary can be taken as two truncated circumstellar
discs. These discs will then evolve independently, {\em unless}
a circumbinary disc exists and material is able to flow
across the gap (for a review, see Lubow \& Artymowicz 1996).
This process is predicted to occur, though observationally
circumbinary discs do not appear to be common (Mathieu 1996).

\subsection{Discs in systems of varying separation}

The simplest effect of binary companions on disc evolution
might be expected to be a systematic variation of disc 
lifetime with binary separation. The reduced viscous time 
$t_\nu \simeq R_{\rm out}^2 / \nu_{\rm out}$ at the outer edge 
should lead to faster disc evolution in closer systems, and 
a higher fraction of WTTS as compared to CTTS. This effect 
has been searched for (Ghez, Neugebauer \& Matthew 1993; 
Simon \& Prato 1995), but although there are systems 
such as DI Tauri (Meyer et al. 1997) that seem to show
the expected effect there appears to be
no fully convincing evidence for a general trend of CTTS fraction 
with separation. Extension of the observations to smaller
separation binaries should be able to determine if this 
implies ubiquitous replenishment of circumstellar discs
from a circumbinary reservoir, as is already implied 
from the mere existence of discs in the very closest 
(spectroscopic) binaries.

\subsection{Discs in systems with unequal mass ratios}

\begin{figure}
 \psfig{figure=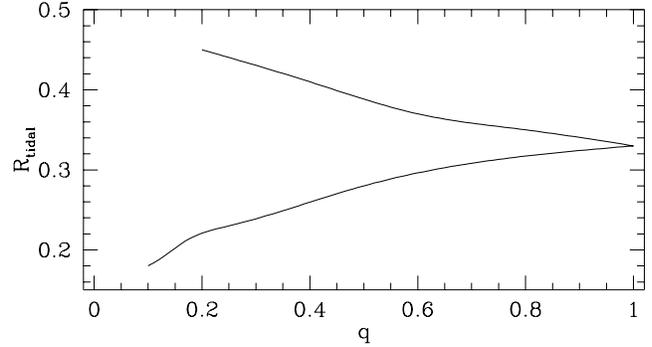,width=3.5truein,height=3.5truein}
 \vspace{-1.5truein}
 \caption{The assumed outer radius of the accretion disc of the
 primary (upper curve) and secondary (lower curve) as a 
 fraction of the binary separation, for various mass ratios
 $q$. Adapted from Papaloizou \& Pringle (1977).}
 \label{figure6}
\end{figure} 

For binaries with unequal mass ratios, we can also ask 
whether the different truncation radii of discs surrounding 
the primary and secondary will lead to observable differential 
disc evolution. Figure \ref{figure6} shows the predicted outer radii
of tidally truncated discs in a binary system as a function
of the mass ratio $q=M_{\rm secondary} / M_{\rm primary}$
(results taken from Papaloizou \& Pringle 1977; see also
Paczy\'nski 1977). We take the `primary' to be the more
massive star, so that $q \leq 1$.
For binaries with $q \sim 0.5$, the
ratio of the outer radii of their accretion discs is
expected to be $\sim 1.5$, rising to $2-3$ for more extreme
systems with $q \sim 0.1$. This disparity in radii will
affect the evolutionary timescales of the two discs,
while the epoch at which the transition from CTTS to
WTTS status occurs will also depend on the disc initial
conditions. 

If the discs survive for long enough
the accelerated evolution of the secondary's smaller
disc compared to the primary's 
will eventually overwhelm any differences
in the initial conditions of the two discs. Whatever the
initial ratios of surface density, disc mass and accretion
rate, at late enough epochs we thus expect that the secondary
should display the weaker disc of the pair. Our 
numerical results aim to quantify when that switchover
should occur, and how the models compare to the growing
body of observations of close binary systems.

\begin{table}
\caption{Summary of the initial conditions for the binary calculations}
\begin{tabular}{lcccc}
      & $M_* / M_\odot$ & $R_{\rm out} / {\rm a.u.}$ & $\dot{M}_{\rm init} /
M_\odot \ {\rm yr}^{-1}$ & $M_{\rm disc} / M_\odot$ \\ \hline

$q=0.44$ & 0.9885 & 20 & $3 \times 10^{-7}$ & 0.076 \\
         & 0.9885 & 20 & $3 \times 10^{-7}$ & 0.020 \\
         & 0.4375 & 13.5 & $3 \times 10^{-7}$ & 0.036 \\
         & 0.4375 & 13.5 & $1.3 \times 10^{-6}$ & 0.096 \\
&&&&\\

$q=0.19$ & 0.9885 & 22.5 & $3 \times 10^{-7}$ & 0.092 \\
         & 0.1901 & 11   & $3 \times 10^{-7}$ & 0.023 \\
         & 0.1901 & 11   & $1.4 \times 10^{-6}$ & 0.063 \\
&&&&\\

$q=0.1$ & 0.9885 & 23.5 & $3 \times 10^{-7}$ & 0.10 \\
	& 0.10   & 9    & $3 \times 10^{-7}$ & 0.015 \\
	& 0.10   & 9    & $1.4 \times 10^{-6}$ & 0.042 \\
\hline	          

\label{binary_table}
\end{tabular}
\end{table}

Table \ref{binary_table} summarises the main parameters for 
the numerical runs. In all cases we consider a binary with 
separation 50 a.u., take $\alpha = 5 \times 10^{-3}$, and impose
a $v_{R} = 0$ boundary condition at $R_{\rm out}$. The smaller 
value of $\alpha$ gives similar disc mass fractions 
as for the models in Section 3. 600 radial
grid points are used. 

The initial conditions for the primary are taken as a steady-state
disc with an accretion rate of $\dot{M} = 3 \times 10^{-7} \ M_\odot
{\rm yr}^{-1}$, that fills its Roche lobe out to $R_{\rm out}$. For
the $q=0.44$ case a weakened initial disc was also run, with
the same initial accretion rate but with the radial extent of the disc
truncated such that $M_{\rm disc}$ was reduced by a factor of
approximately four. 

For the discs surrounding the secondaries, our standard case
commences with an accretion rate equal to that of the primary.
These discs are initially a factor 2-6 times less massive than
the corresponding primary disc. However, cognisant of the 
fact that in some formation scenarios the secondary disc
might accrete more infalling matter than the primary (Bate \&
Bonnell 1997), we also consider initially higher accretion
rates through the secondary's disc. These are chosen such
that the Q parameter, 
\begin{equation}
 Q= {{c_s \Omega} \over {\pi G \Sigma}}
\label{Q_def}
\end{equation}
equals unity at $R_{\rm out}$, and represent the maximum
plausible accretion rates through steady discs described by the
vertically averaged equations. For $Q \ale 1$ gravitational
instability sets in (Laughlin \& Bodenheimer 1994), and leads
to a rapid redistribution of disc angular momentum. A more
massive disc would thus be expected to relax to a gravitationally
stable configuration on a timescale that is short compared to the 
viscous timescales over which we compute the evolution. These
high $\dot{M}$ secondary discs have surface densities that 
are initially considerably higher than for the primary, and
masses that are comparable, or (in the $q=0.44$ case) slightly
larger.

\begin{figure}
 \psfig{figure=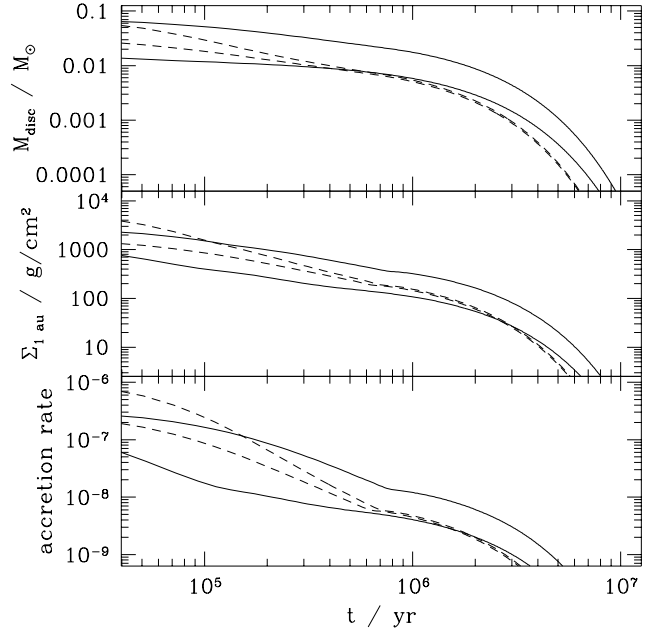,width=3.5truein,height=3.5truein}
 \vspace{-0.0truein}
 \caption{The decline of the disc mass, surface density at 1 a.u., and
 	  accretion rate (in $M_\odot {\rm yr}^{-1}$), 
          for the $q=0.44$ calculation. The solid curves show the
          primary disc for the standard (upper curves) and low mass (lower
          curve) disc initial conditions; the dashed lines show the secondary disc
          for high and low initial accretion rates. Here we have assumed 
          a binary separation of 50 a.u., and taken $\alpha = 5 \times 10^{-3}$.}
 \label{figure7}
\end{figure}

Figure \ref{figure7} shows in detail the evolution of
the discs for the runs intended to mimic a $q=0.44$ binary
system. Three possible indicators of the strength of the
circumstellar discs are plotted, the total disc mass, the
surface density at a radius of 1 a.u., and
the accretion rate, which is expected to correlate well with
measures of activity based on H$\alpha$ and UV flux. 

For the standard initial conditions of the primary disc, and for
all three measures of disc strength, the faster evolution
of the smaller disc around the secondary is sufficient to
weaken that disc below that of the primary in
a relatively short time $\sim 10^6$ yr or less. The initial
conditions matter most for measures based on the accretion
rate, but even here the primary disc is substantially 
stronger than that of the secondary at accretion rates
well above the typical detectable threshold.

If the disc of the primary is initially substantially smaller
than the tidal truncation radius then a much longer
period is required before the secondary's disc becomes
weaker than that of the primary. In particular, the
accretion rate through the primary disc remains below that of 
the models for the secondary down to $\dot{M} \approx 3 \times
10^{-9} \ M_\odot {\rm yr}^{-1}$, which would be hard to 
measure unambiguously observationally. Hence if the initial
conditions were as biased in favour of a strong secondary 
disc as in this calculation they would overwhelm the
faster evolution of the secondary's disc for all 
observations dependent on the disc accretion rate.

\subsection{Variation with mass ratio}

\begin{figure*}
 \psfig{figure=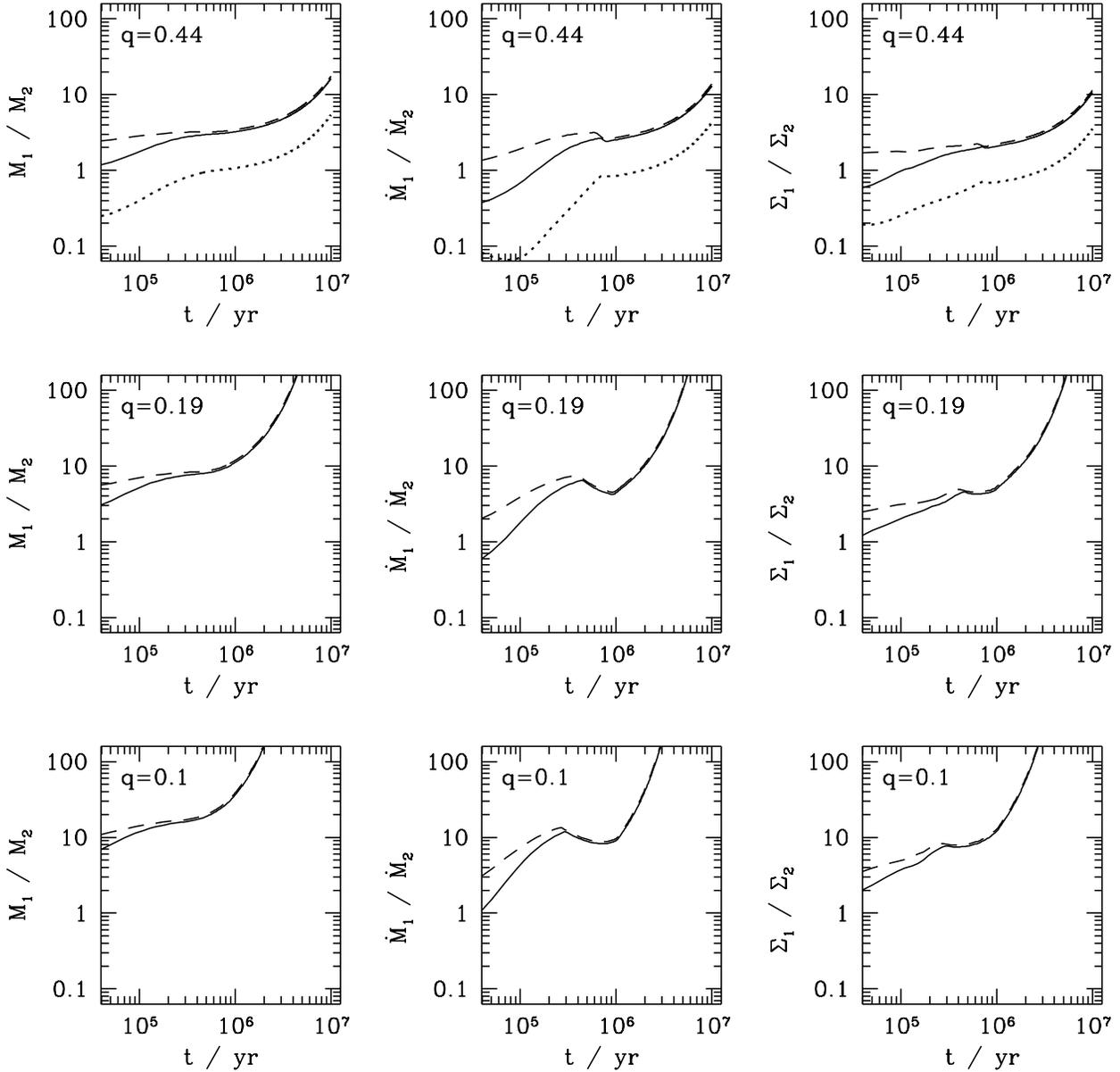,width=7.0truein,height=7.0truein}
 \vspace{-0.3truein}
 \caption{The relative strength of the primary and secondary 
          discs, as a function of time, for varying mass ratios $q$.
          Disc `strength' is measured by disc mass (left panels), 
          accretion rate onto the star (centre panels), and surface 
          density at a radius of 1 a.u. (right panels). The solid 
          lines show the ratio of the strength of the primary to 
          secondary discs for models where the secondary disc 
          initially has a high accretion rate. The dashed lines
          show the case where the two stars have initially identical 
          accretion rates. For a mass ratio $q=0.44$ we also show
          with the dotted line the case where the primary disc is
          initially of low mass, as described in the text.}
 \label{figure8}
\end{figure*}

Figure \ref{figure8} summarises the results of the
runs in Table \ref{binary_table}. For each mass ratio
and indicator of disc strength (disc mass, accretion rate, 
and surface density at an arbitrary radius of 1 a.u.) we
plot the relative strength of the primary and secondary 
discs as a function of time. As in Section 3 we caution 
that the absolute time scaling of the models is somewhat 
arbitrary -- $\alpha$ has been chosen so as to give 
plausible disc lifetimes in the Myr range.

As we have already discussed, for $q=0.44$ the evolutionary 
timescales of the two discs are sufficiently similar that 
the initial conditions are important through to late times. 
For reasonable choices, there is a lengthy period when the 
primary and secondary discs have at least comparable strength. 
Observations of such binaries thus potentially retain 
information on the formation histories of the discs.

Few binaries appear to have very extreme mass ratios. However 
studies of those systems can provide a stringent test of the 
evolution of the disc, since for $q \ale 0.2 - 0.3$ the large 
disparity in evolutionary timescales is sufficient to rapidly 
overwhelm variations in the initial disc masses and radii. From 
Fig.~\ref{figure8} it can be seen that for the two most extreme 
mass ratios essentially all binaries are predicted to have a stronger
circumprimary disc than circumsecondary. This applies for all 
measures of disc strength, though the disc mass is always the 
most robust measure. 

The results can also be used to estimate the predicted fraction 
of `mixed' systems, i.e. the number of binaries pairing 
a WTTS with a CTTS as a fraction of the total number of binaries
containing at least one CTTS. To do this we assume that the 
transition from CTTS to WTTS occurs at some fixed value of the 
disc mass (or accretion rate or surface density) for both 
primary and secondary, and use the numerical models to provide 
the time spent as a doubly strong (CTTS + CTTS) and as a mixed 
(CTTS + WTTS) system. 

For the $q=0.44$ run, the results depend strongly on when 
the epoch of disc dispersal occurs. If the disc is lost 
relatively early on, when only a few viscous times have 
elapsed, then the predicted fraction of mixed systems 
depends almost entirely on the uncertain initial 
conditions. 

For lower-mass-ratio systems, where the disparity in
evolutionary timescales is greater, the influence of
the initial conditions is less important. 
For $q=0.19$ the time spent in the `mixed'
state is expected to exceed that spent as a doubly strong
system, while for $q=0.1$ the ratio of times is $\sim 5$
for all measures of activity (i.e. the fraction of mixed
systems amongst all binaries containing at least one CTTS would 
be $\sim 0.8$). This applies even if the discs are destroyed at 
an accretion rate of order $10^{-8} \ M_\odot {\rm yr}^{-1}$. The
observation of even a few systems where the mass ratio 
was this extreme can thus provide a stringent test --
the standard models presented here are incompatible both
with a low fraction of mixed (CTTS, WTTS) systems, and
with systems where the primary is weaker than the secondary 
at any except the highest accretion rates. Observations
to the contrary would be good evidence for accretion from
a circumbinary disc.  

\subsection{Discussion}

In this Section we have investigated the evolution of
discs in close binary systems. We find that the secondary's
disc should always become weaker than that surrounding the primary
at a sufficiently late epoch, but that the timing of the
switchover is a strong function of the mass ratio of the binary.
For mass ratios of $\age 0.5$, a secondary that has an initially
higher accretion rate through its disc might remain stronger
than the primary until the disc is cleared. For more extreme
mass ratios, we expect the faster evolution of the smaller
secondary disc to dominate over the influence of the initial  
conditions. At late times the secondary should display 
weaker disc signatures than the primary, and there should
be a relatively high fraction of mixed T Tauri binaries
pairing a CTTS primary with a WTTS secondary.

Current observations appear to be broadly consistent
with this picture. Brandner \& Zinnecker (1997) studied
a sample of binaries in the 90-250 a.u. range, finding
3 systems (out of 12 containing a CTTS) where a CTTS
secondary was paired with a WTTS primary. In this
sample, all the measured mass ratios were $q > 0.5$,
and the analysis was based on H$\alpha$ flux. This
sample thus falls into the category where we would predict
that the initial conditions of the discs formed around
the stars should be important, and strong accretion activity
among the secondaries is not unexpected.

The photometric properties of close T Tauri binaries
have been studied by Prato \& Simon (1997), and Ghez,
White \& Simon (1997). These studies show that mixed
(CTTS, WTTS) binaries appear to be rare -- {\em none}
were found in the sample of 12 systems examined
by Prato \& Simon. This is probably marginally
consistent with our models, at least if discs in these
binary systems are cleared at a relatively high 
accretion rate. However, it is clear
that a similar observational result for systems with
an extreme mass ratio would be inconsistent with these
independent evolution models, and would point either to 
common replenishment of the discs from a 
circumbinary disc, a much weaker dependence of $t_\nu$ on 
$R$ than that assumed here, or a coordinated mechanism for
their common destruction. 

\section{Conclusions}

In this paper we have presented models for the evolution
of circumstellar discs around T Tauri stars. The
models combine a pre-main-sequence stellar evolution 
track with a model for viscous disc evolution. Over 
the estimated 1-10 Myr lifetime of T Tauri discs (Strom 1995)
there are large changes in both the stellar luminosity 
and the disc accretion rate (Hartmann et al. 1998), 
making the inclusion of both components essential in 
an evolutionary model.

Comparison of our theoretical tracks with the distribution 
of infra-red colours of T Tauri stars in Taurus-Auriga 
(Kenyon \& Hartmann 1995) yields two main conclusions. 
Most securely, in all models the transition in the colour-colour
plane between systems looking like Classical T Tauri stars, 
and those appearing as Weak-lined systems, is slow, and occurs 
on the viscous timescale of the outer disc. This occurs even if 
a dynamically significant stellar magnetic field disrupts the 
inner disc. However observationally this transition {\em must 
occur very rapidly}, since there are few systems with properties 
intermediate between CTTS and WTTS. Non-viscous
processes must be the agents of this transition. 

A more detailed comparison of the models to the data 
provides some constraints on when disc dispersal must occur.
The presence of dynamically significant stellar magnetic
leads to a distinctive evolutionary track of stars
in the (K-L)-(K-N) plane. As the accretion rate declines,
the (K-L) colour drops rapidly to close to bare photospheric levels 
while the system is still red in (K-N). No systems with this 
combination of colours are observed, which implies that the 
models can only be consistent with the observations if 
purely reprocessing discs are rare. We speculate that
changes in the disc structure at the epoch when the disc 
goes from being active (heating dominated by viscous processes), 
to passively reprocessing stellar radiation, could be 
responsible for initiating the disc's rapid demise. 

Applying the models to close binary systems, we find that
the observed lack of systems pairing CTTS with WTTS is
consistent with disc destruction occurring at a similar
accretion rate to that inferred from the single star
models. The models suggest that the frequency of mixed
T Tauri binaries should increase greatly at extreme 
mass ratios. Observations of such systems, and studies
that probe the lifetime of the CTTS phase as a function
of separation, are a good indicator of T Tauri disc
evolution and the ubiquity of accretion from circumbinary
discs. Conversely, observations of discs in binaries with more
nearly equal mass components probe the formation history of those
discs, especially if observed at early epochs.

For the subsample of systems with measured photometric 
periods, we find no correlation between the infra-red
colours and the photometric period, which we assume is probably
a good tracer of the stellar rotation rate. If the magnetospheric
radius was approximately equal to the corotation radius, or
a fixed large fraction of it, a correlation of (K-L)
with $P_*$ would be expected. The lack of such a correlation, 
along with suggestions that the magnetosphere lies well 
inside the radius of corotation (Kenyon, Yi \& Hartmann 1996; 
Meyer, Calvet \& Hillenbrand 1997), pose potentially serious
problems for models that seek to explain the slow rotation 
of Classical T Tauri stars as a result of magnetic linkage
between the star and its disc.

\section*{Acknowledgements}
We thank Andrea Ghez
and Scott Kenyon for very useful discussions, and 
Alexander Men\' shchikov for
supplying the dust opacity used in this work in computer
readable form. The hospitality of the National Astronomical 
Observatory and the Isaac Newton Institute for Mathematical 
Sciences, where parts of this work were completed, is 
gratefully acknowledged.

\end{document}